# Developing a new biophysical tool to combine magneto-optical tweezers with super-resolution fluorescence microscopy

Zhaokun Zhou[a, 1], Helen Miller[a, 2], Adam J. M. Wollman[a, 3] and Mark C. Leake[a, *]

[a] Biological Physical Sciences Institute (BPSI), Departments of Physics and Biology, University of York, York YO10 5DD, United Kingdom

[1] E-Mail: zz840@york.ac.uk;
[2] E-Mail: hlm531@york.ac.uk;
[3] E-Mail: adam.wollman@york.ac.uk;

[*] Author to whom correspondence should be addressed; E-Mail: mark.leake@york.ac.uk; Tel.: +44 (0)1904 322697 (Physics office); +44 (0)1904 328566 (Biology office).

**Abstract:** We present a novel experimental setup in which magnetic and optical tweezers are combined for torque and force transduction onto single filamentous molecules in a transverse configuration to allow simultaneous mechanical measurement and manipulation. Previously we have developed a super-resolution imaging module which in conjunction with advanced imaging techniques such as Blinking assisted Localisation Microscopy (BaLM) achieves localisation precision of single fluorescent dye molecules bound to DNA of ~30 nm along the contour of the molecule; our work here describes developments in producing a system which combines tweezing and super-resolution fluorescence imaging. The instrument also features an acousto-optic deflector that temporally divides the laser beam to form multiple traps for high throughput statistics collection. Our motivation for developing the new tool is to enable direct observation of detailed molecular topological transformation and protein binding event localisation in a stretching/twisting mechanical assay that previously could hitherto only be deduced indirectly from the end-to-end length variation of DNA. Our approach is simple and robust enough for reproduction in the lab without the requirement of precise hardware engineering, yet is capable of unveiling the elastic and dynamic properties of filamentous molecules that have been hidden using traditional tools.

**Keywords:** optical tweezers; magnetic tweezers; super-resolution microscopy

# 1. Introduction

Molecular force and torque play fundamental roles in biology. They drive many mechanical and chemical processes on a molecular level. Being capable of applying and measuring force and/or torque, magnetic and optical tweezers have been hugely successful at studying molecular structures and dynamics, for example, in probing the action of topoisomerases, FtsK and $F_0F_1$ ATPase [1-4] and in investigating DNA conformation and elasticity.

While it is true that optical tweezers (OT), also referred to as laser tweezers, as well as magnetic tweezers (MT), are capable of simultaneous transduction of both force and torque, using either technique for the transduction of both has serious limitations. Optical tweezers are ideal for force transduction investigations [2] but they are limited in transducing torque. Laguerre-Gaussian beams contain orbital angular momentum, which can impart angular momentum on probe particles leading to their rotation. This constant-torque, rather than constant-rotation, transduction means it is difficult to precisely control the speed and number of rotations of the trapped particle. Also the transparency of the probing particle has to be low enough to absorb enough photons to rotate and be sufficiently high for tweezing [5]. Alternatively, linearly polarised light does controllably rotate a birefringent probe a defined number of turns, but the requirement that the extraordinary axis is perpendicular to the trapping beam is technically challenging and often demands very precise nanofabrication of non-spherical probes [6]. Another limitation is the difficulty in multiplicity. Multiple traps via time-sharing does not allow torque to be applied continuously, and via spatial separation [7, 8], multiplies the sophistication of the optical system. Holographic optical tweezers specialise in multiplicity but lack the ability to individually rotate the beads [9].

Similarly, while magnetic tweezers are excellent at applying torque [1], simultaneous force application has its complications. The magnetic bead used in MT can be modelled as a magnetic dipole, which moves along the magnetic field gradient towards a local B-field maximum. But a B-field maximum never resides in free space. So unlike laser tweezers, magnetic fields cannot create a stable stationary equilibrium configuration to trap the bead (Earnshaw's Theorem). Special geometries have to be used to circumvent this – with the B-field pulling the bead in one direction while a separate force pulls in the other to restrict the bead position. The second force can be from flow pressure [10], or from the biological molecule itself [11]. Alternatively, sophisticated electromagnets [12] and micro-electromagnets [13, 14] with multiple coils arranged around the bead can keep the bead quasi-stationary with positive feedback. Besides not being possible to create static field gradients in the imaging plane, the inability to decouple force and torque has limited experimental investigations.

For DNA stretching experiments, forces ranging from a few pN to tens of pN are usual, with B-field gradient capable of applying such forces inevitably requiring B-field strengths that can exert torques at least three orders of magnitude above biologically relevant values and above values measurable via noise power spectrum methods. Clever geometries involving the combination of a cylindrically symmetric torque-less magnet to pull the bead and a small side magnet to apply force have been devised [15]. Similar designs with the side magnet replaced by electromagnets can reduce mechanical vibration and increase control precision [16]. Also, rotation faster than the angular response bandwidth of the bead has been utilised to rotate the bead at a speed non-linearly dependent on the B-field rotation frequency [17]. However, these measures all make calibration and use of the devices challenging.

In biological reactions where both force and torque are involved, it is intuitive and logical to combine optical and magnetic tweezers to achieve simultaneous and independent force and torque transduction and measurement. Besides avoiding the pitfalls of using either device on its own, the combination also has the advantage of high temporal resolution afforded by the quadrant photodiode (QPD) back focal plane (BFP) interferometry detection of a bead's 3D position and rotation, the lack of which has restricted magnetic tweezers to low resolution measurements. In magnetic-tweezers-only setups, camera video imaging is traditionally used to track the bead. Despite the fact that typical maximum camera frame rates have improved dramatically over the past decade from merely 'video rate', at high frame rates (>10 kHz) the photon count per pixel per frame drops dramatically, increasing the signal-to-noise ratio and limiting the spatial resolution to worse than 4 nm [18], whereas spatial resolutions as small as 0.1 nm have been reported with QPD interferometry detection [19]. Besides, the 10 kHz frame rate is still a far cry from the 200 kHz bandwidth readily achievable with QPD detection.

Another bonus of torque-only magnetic tweezers applications is the dramatic reduction of the B-field strength. Typical molecular forces and torques relevant in experiments are on the order of one to tens of pN and one to tens of pN·nm respectively. For example, the B-S DNA transition happens at 65 pN [20] and P-B transition at 34 pN·nm [21]. In a typical electromagnet design using a 2.8 μm superparamagnetic bead (M280, Dynal Inc., Lake Success, NY), a B-field gradient of around 10 T·m$^{-1}$ can exert a maximum force of 0.1 pN, which is less than that used in DNA twisting experiments. This gradient corresponds to a B-field around 0.1 T. Torques, however, require no field gradient and a strength of 0.1 T will generate up to $10^7$ pN·nm [22] on the same bead!

This reduction in B-field no longer necessitates the use of permanent magnets, the rotation of which is limited to less than 30 rev/sec and which introduces vibrational noise into the system. In the case of electromagnets, designs can avoid the inclusion of pole pieces, extension structures that guide the B-field to the bead and connective caps that close the field lines. All of these contribute to hysteresis and to varying degrees of eddy currents which both dissipate energy as heat. Low B-fields also reduce the current to such an extent that Ohmic heating from the coil itself is negligible so cooling the system is unnecessary. This frees up precious space on the setup, the lack of which in a commercial microscope with a piezo nanostage control always handicaps the exploration of design features.

Optical and magnetic traps have long been combined in cold atom research in condensed matter physics, to trap and cool individual atoms. However, the same combination in molecular biology has only attracted limited attention [23-28] and has yet to achieve its full potential. The earliest attempt arranged four coils above and four below the sample stage. Inserted into each coil is a pole piece attached to a tip-pole expansion guiding the B-field to the sample [23]. The pole pieces cause hysteresis and require careful calibration. Also the soft iron used to support the structure and close the B-field lines allows the generation of eddy currents that impose uncharacterised impedance on the driving current, resulting in heating at high field frequencies. The tip-pole expansion severely limits the space available for the sample chamber. Claudet and Bednar improved the design by restricting rotation to the one dimension required for stretching experiments using a near-Helmholtz coil configuration [25]. However, their horizontal coils prevented the use of oil immersion condensers, limiting the resolution of BFP interferometry bead tracking. Also the small coils make field uniformity more vulnerable to misalignments in coil spooling, which is inevitable due to the non-zero thickness of the wires.

Experiments with molecular force/torque manipulation and measurement capabilities will be brought to a new level with single-molecule fluorescence imaging. Pre-fluorescence imaging methods relied on the imaging of the probe itself, such as tracking the bead position to deduce 'plectoneme' formation (a spatially localized supercoiled structure) in a DNA twisting experiment. Not only are those methods incapable of observing weak intermolecular reactions that do not result in measurable changes in molecular length, but also they provide no information on the location of events. Super-resolution imaging with fluorophores tagged onto the macromolecule directly monitors macromolecular conformational changes and inter-molecular reactions.

Super-resolution fluorescence microscopy pushes the resolution of single-molecule imaging [29, 30], with techniques such as STORM/PALM [31, 32] and BaLM [33]. Our lab recently achieved lateral localisation precision of 30-40 nm with video rate imaging of YOYO-1 dye stained DNA [34]. Through many cycles of image capturing, fluorophore bleaching, and random simultaneous recovery, we were able to obtain fluorescence images of discrete separated emission signals. We used Matlab code [35] to localise the emissions from a series of images and reconstruct DNA molecules. Imaging was also performed on DNA-bead complexes as proof-of-principle to show the compatibility with magneto-optical tweezers manipulation. We were also able to observe dynamic topological changes in real-time, paving the way for real-time DNA stretching/twisting type imaging.

**2. The instrument**

The design is driven by the need for versatile biological applications mainly involving probing and imaging dynamic single-molecule topology of DNA, and protein machines that manipulate DNA topology. The filamentous molecule needs to be in a transverse orientation for contour-wise imaging. This defines the rotational axis. The space available from the commercial microscope and the piezo stage rules out the option of permanent magnets, which are attached to rotors. Electromagnets with multiple cores are also ruled out for the same reason. We have reduced the system to the minimum configuration with four coils, without pole pieces, that is capable of rotation in one horizontal dimension but at the same time compact enough to fit into a nano positioning stage whilst allowing space for optical tweezers.

The centre of the field of view needs to coincide with the centre of the B-field for maximum uniformity and needs to be stationary relative to the B-field. Since the objective lens turret cannot mechanically sustain the weight of the magnetic tweezers without bending, a platform was built directly into the microscope body for the MT to be mounted (see Supplementary Information, Figure 2).

Since the B-field will only be responsible for torque, it is kept as uniform as possible. Two pairs of Helmholtz coils placed at right angles are known to be the simplest configuration for uniform field generation but the space available around the sample stage dictates that the coil pair is slightly further apart than that of a Helmholtz configuration (see Supplementary Information Figure 1 for details of the coil holder design). This introduces small non-uniformity into the field, which is discussed later. The coil support structure is CAD designed and 3D printed (printer: Object30, material: VeroWhitePlus RGD835) to optimise space use and to maximise field uniformity. SWG 20 enamelled copper wires (05-0240, Rapid Electronics Ltd.) are manually wound onto the spools. The left and right spools have 95 turns and the top and bottom spools have 100 turns to fill up spool space. The diameter of the electromagnetic wires does not significantly affect the field strength or heat production (see Supplementary Information, Heat dissipation *vs* wire thickness calculation). But too few turns requires a high current that only an expensive specialised current source/sink could supply. Furthermore, overly thick wires adversely affect winding quality.

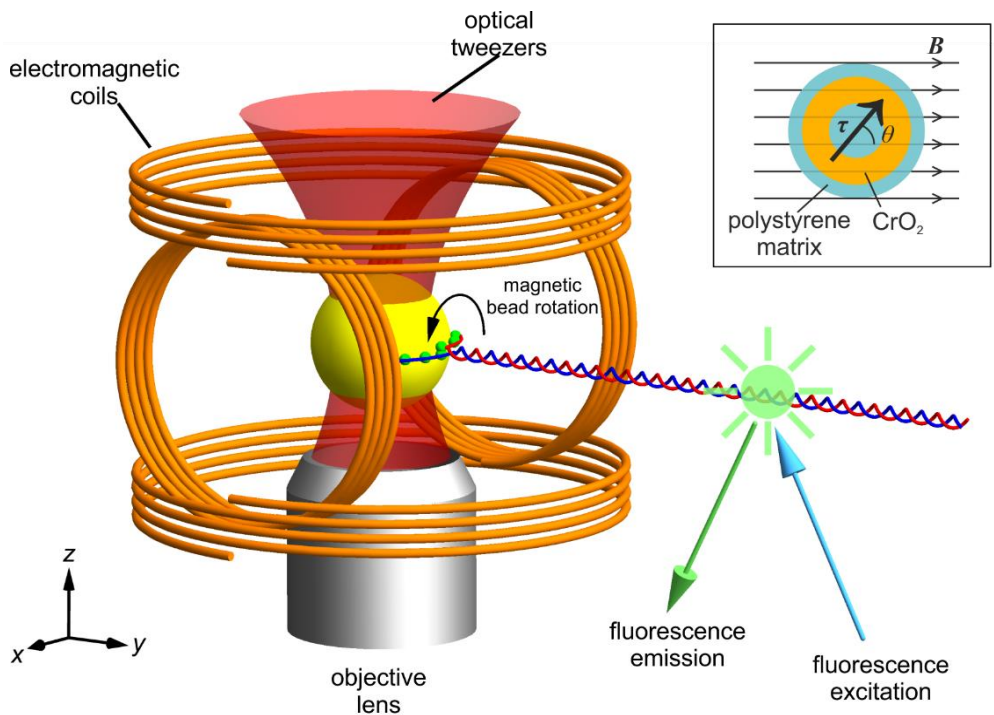

**Figure 1.** Schematic of our magneto-optical tweezers combined with fluorescence microscopy and the ferromagnetic bead in a B-field. The magnetic tweezers are comprised of two pairs of parallel coaxial coils arranged at right angles to each other in the *x-z* plane. This makes the *y* axis the rotational axis. The optical trapping laser beam enters from below the objective lens. The DNA molecule is pulled along the *y* axis and tethered to a ferromagnetic bead at multiple points (green dots) to lock rotation. Fluorophores are bound along the length of the DNA for super-resolution imaging of DNA topology. Note that the bead and DNA are shown not to scale for clearer representation. The inset shows the schematic of the cross-section of a typical ferromagnetic bead. The beads we use have an average diameter of 2.10 μm. Note the magnetisation is at an angle $\theta$ from the background field. This angular displacement gives rise to a torque applied onto the bead $\boldsymbol{\tau} = \boldsymbol{B} \times \boldsymbol{m} = Bm\sin(\theta)$.

The signal generation is controlled via our custom LabVIEW software (National Instruments Corp., Sunnyvale, CA) and executed in an analogue output device (NI 9263 and NI cDAQ-9174). Then we use a car audio amplifier (Pioneer GM-D8604 1200W 4 Channel Class D Car Amplifier) to raise the current to the required amplitude. In our case the current is sinusoidal without any DC offset. This low cost approach proves to be ideal for our application. The root-mean-square noise introduced in the amplification circuit manifests as noise in the B-field. Due to the low magnitude of the field, a Gauss meter was not stable enough to measure the field. We instead calibrated the field by monitoring the rotation of the bead itself.

The optical tweezers are built around an inverted microscope (Nikon Eclipse Ti-S, Nikon Instruments Inc.) with two laser sources, two QPDs for bead tracking and two cameras for fluorescence imaging. A white-light laser (Fianium SC-400-4, Fianium Ltd.) whose spectrum covers ~480 – 2400 nm provides both the excitation light for fluorescence imaging and the tracking beam for interferometry. A near-infrared laser (1064 nm, max output 4W, Elforlight L3000-1064) provides the trapping beam. The beam is expanded 10-fold to slightly under-fill [36] the entrance pupil of the objective (100x, NA 1.45, oil immersion, model no. MRD01095, Nikon Instruments Inc.) so as to achieve a stiff trapping spring constant. An acousto-optic deflector (part no. DTD-274HD6M, IntraAction), controlled by RF synthesiser (part no. DVE-120, IntraAction) and amplifier (part no. DPA-502D, IntraAction), is used to create multiple time-shared beams with high frequency beam steering for each trap to compensate for noise and drift. After the objective, the beam focuses to a diffraction limited spot before diverging again. An oil immersion condenser (NA 1.4, Nikon Instruments Inc.) re-collimates the beam and forms an interference pattern due to the scattered beam (by the bead) and the unimpeded beam at its back focal plane. An imaging lens projects this pattern onto a QPD (QP50-6-18u-SD2, First Sensor) placed at the conjugate plane to the condenser BFP to detect the interference signal, which is fed to an analogue input device (NI 9222 and

NI cDAQ-9174) for rapid data analysis. A second QPD allows separate monitoring of the high frequency noise in the flow-cell for stabilisation with the piezo nanostage.

The fluorescence excitation path utilises the visible spectrum of the white-light laser. Two excitation channels were created using a bespoke two channel tuneable colour splitter consisting of a 552 nm dichroic and two pairs of linear filters (Delta Optical Thin Film) on adjustable mechanical mounts, allowing the centre wavelength and bandwidth to be set. A beam width of FWHM 57.3μm (widefield) or 10.7 μm (narrowfield/Slimfield) [37, 38] at the sample plane can be chosen depending on application, and either beam width can also be used in Total Internal Reflection Fluorescence (TIRF) and oblique epifluorescence/HILO mode [39]. Fluorescence emission signals are separated by a dichroic into two channels (currently at 580 nm) and separately imaged onto two EMCCD cameras (iXon Ultra 897, Andor Technology Ltd) for bi-chromatic fluorescence imaging.

The laboratory housing of our device is air conditioned (MFZ-KA50VA, Mitsubishi Electric) to contain temperature fluctuations within ±0.1 °C. All equipment with moving parts such as fans are mounted away from the optical table in the same room. All optical components and the magnetic tweezers are mounted on an air-cushioned surface (PTQ51504, Thorlabs Inc.) to minimise acoustic and mechanical noise and are fully contained in aluminium boxes or tubes (except the body of the actual microscope) so no beam is exposed. This reduces air-flow and dust that potentially compromises laser profile and coherence.

The flow chamber is formed with a coverslip taped on two opposite sides onto a microscope slide. The chamber is then passivated with BSA to prevent beads from sticking to the surface. Then the bead samples are pipetted into the flow chamber by capillary effects.

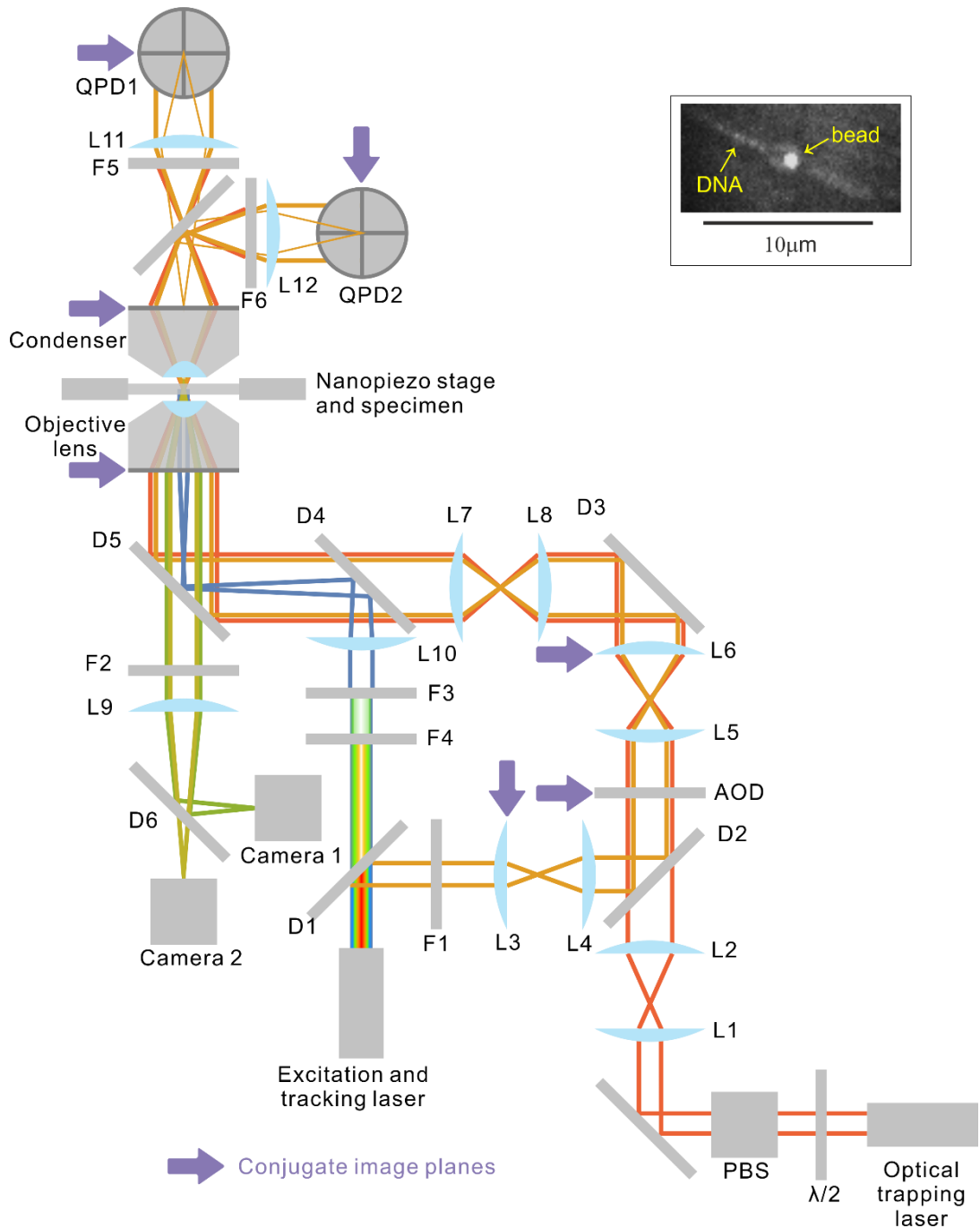

**Figure 2.** Schematic of the optical system (magnetic tweezers not shown) and a fluorescence micrograph of bead-tethered DNA (inset). Immediately next to the optical trapping laser, a half-wave plate (WPH05M-1064, Thorlabs Inc.) and polarisation beam splitter (PBS123, Thorlabs Inc.) attenuate the output power. Lens pair (L1, L2) expands and collimates the trapping laser (Elforlight L3000-1064, Nd:YAG) beam before the beam enters the AOD. Lens pair (L5, L6) further expands the beam to slightly under-fill [36] the entrance pupil of the objective. Lens pair (L7, L8) is a 1:1 telescope that images the back focal plane of the objective onto L6. L6 is mounted on a 3-axis translational stage to enable manual adjustment of both the trapping and tracking laser beams. The tracking

and fluorescence excitation beams are extracted from the same laser, which emits a continuous spectrum from wavelength ~480 nm to ~2400 nm. The spectrum is first split by a hot mirror (D1, cut off frequency 700 nm, M254H45, Thorlabs Inc.) into the visible part and near infrared part. For simplicity, the two channel colour splitter that divides the visible part into two channels is not shown. The long wavelength part that reflects off D1 has an 830 nm line singled out with a notch filter (F1, FF01-830/2-25, Semrock Inc.) for position tracking of the trapped bead. The tracking beam is first expanded by the lens pair (L3, L4) then coupled into the trapping beam via a longpass dichroic (D2, FF875-Di01-25x36, Semrock Inc.) just before the AOD so that both the trapping and the tracking beams are modulated by the AOD. Lens L3 is mounted on an adjustable mount to allow independent steering of the tracking beam. The two beams are then combined with the fluorescence excitation beam via a longpass dichroic (D4, FF775-Di01-25x36, Semrock Inc.) and all three beams reflect off a bandpass filter (D5, FF444/521/608-Di01-22x29) into the objective lens. The beams emerging from the back focal plane of the condenser are then imaged with lenses L11 and L12 onto two quadrant photodiodes (QPDs) for laser interferometry position detection of the trapped beads and for real time stabilisation of the flow-cell [19]. Two line filters F5 and F6 (F1, FF01-830/2-25, Semrock Inc.) prevent the 1064 nm trapping beams from reaching the QPDs. The fluorescent emission is imaged onto two Andor cameras at two wavelength ranges, allowing fluorophores of two colours to be imaged simultaneously. F2 stops any excitation wavelengths from reaching the cameras. All components that are in optically conjugate planes are labelled with purple arrows. The inset on the top right [34] features a fluorescence micrograph of a bead with two YOYO-1 labelled DNA molecules tethered on the opposite sides of the bead.

## 2.1. Calibration

The simplicity and symmetry of the coil geometry allows numerical evaluation of the Biot-Savart law to calculate the field strength. The coils are modelled as concentric rings tightly and uniformly packed. In practice each coil is composed of 10 layers of connected helical coils with alternating handedness. Since the pitch-radius ratios are only 0.028 and 0.024 for the horizontal and the vertical coils and adjacent layers have handedness which cancels, our model is a reasonable approximation. The field at point P($x$, $y$, $z$) due to each ring in each coil is evaluated by the line integral along the ring and the field due to individual rings is summed to find the resultant field. Figure 3 shows the simulated field landscape in the vertical and horizontal plane. The variation of the field in the central 2×2 mm region is no more than 0.02 % per mm, well below the field gradient to exert any measurable magnetic force. This has also been confirmed by measurement.

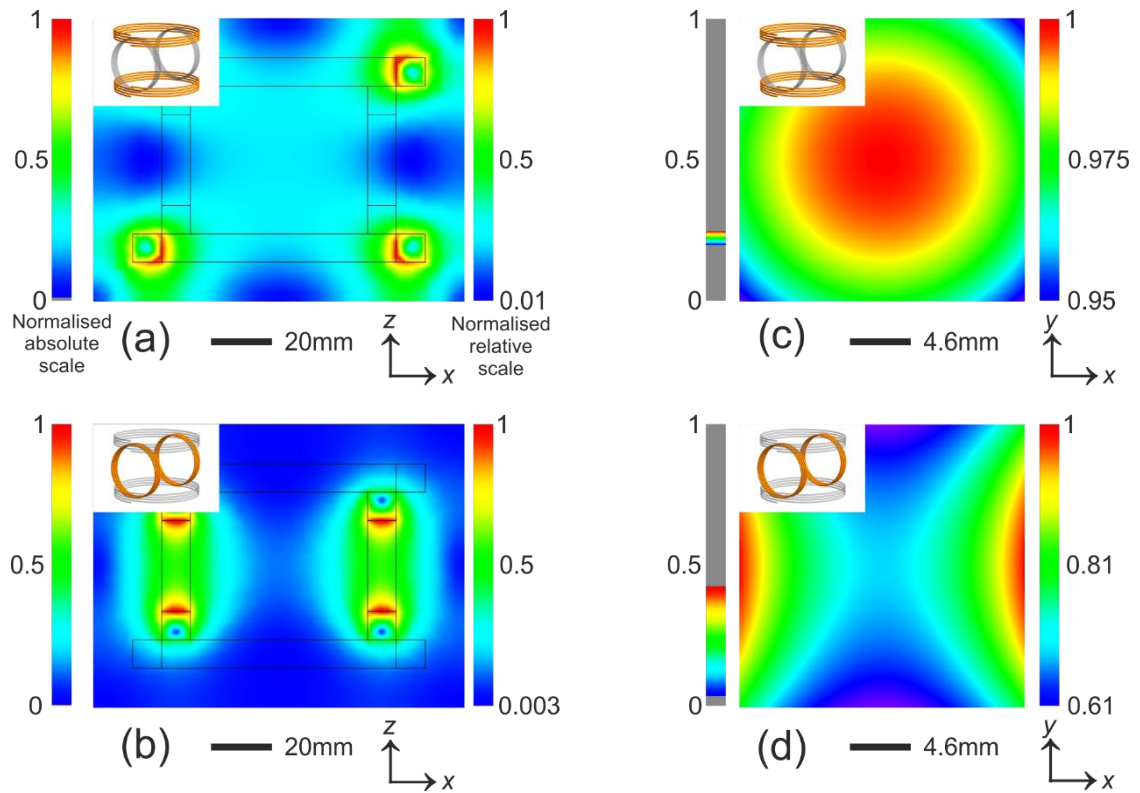

**Figure 3.** Simulation of magnetic field strength. The normalised absolute scale has a maximum value consistent across the entire panel for easy comparison. The normalised relative scale is normalised to the maximum strength in each plot for high contrast. **(a)** and **(b)** show the field in the $y = 0$ plane that incorporates the entire magnetic tweezers. Although the field varies almost from 0 to 1, the central region is relatively uniform. The contour of the coils are sketched. **(c)** and **(d)** zoom into the central region and plot the field in the $z = 0$ plane. The regions are the size of a 22×22 mm coverslip. In **(c)** the field strength varies 0.02 % per mm over a 2×2 mm region, which covers the entire active imaging region and in **(d)** it is 0.02 % per mm over the same region. The force due to this gradient is not measurable. The insets in each plot indicates coils that are turned on (orange) and off (grey).

In a typical MT design, temperature rises pose a significant challenge to experiments as thermal expansion contributes to drift of the bead positions. The total resistive dissipation due to the sinusoidal waves running in our coils is maximally 20 mW (see Supplementary Information, Heat dissipation calculation) and in a 15 minute experiment the increase in temperature we measure is less than 0.1 ℃ on the coils and negligible at the sample plane. The drift caused by thermal expansion in the magnetic tweezers is at most a few tens of nm over a period of 2 hours. This can be corrected by monitoring a reference bead tethered to the flow-cell surface and compensating the position change with the *xyz* nanostage.

The ferromagnetic beads we used are composed of a monodispersed polystyrene substrate coated in a layer of chromium (IV) oxide ($CrO_2$), which is held together by an outer polystyrene layer. Thus they can be modelled as a spherical magnetic shell of uniform thickness and density (see Figure 1, inset). Magnetisation data for 2.10 μm diameter Spherotech ferromagnetic beads (our sample) is not available but is for 4.32 μm diameter [40]. Average remanence for a 4.32 μm diameter bead is $4.35 \times 10^{-13}$ A m$^2$. The manufacturer does not specify how the $CrO_2$ content scales with bead size. Here we assume linear dependence on volume, which gives us average remanence of $5.0 \times 10^{-14}$ A m$^2$. This value is only as good as the assumption we make about the $CrO_2$ content but it provides guidance to experimental design. We model the bead as a magnetic dipole with dipole moment $\boldsymbol{m}$. The torque $\boldsymbol{\tau}$ that the B-field exerts on $\boldsymbol{m}$ is given by $\boldsymbol{\tau} = \boldsymbol{m} \times \boldsymbol{B}$ (see Figure 1, inset). In a DNA twisting experiment, which typically requires a torque of up to $10^2$ pN·nm, the minimum field required is on the order of a few μT. To measure the torque, the imposed B-field needs to be sufficiently small for a measurable angular displacement between the background field and the bead's magnetisation.

The Earth's magnetic field (~50 μT) is negligible in a typical MT assay where the B-field is at least 10 mT so many authors simply ignore it. However, it is clear that here we have to eliminate the effect of this field. A Mu-metal box to shield the region of interest or 3-axis Helmholtz coils to cancel the Earth's field are the usual ways to establish a field-free zone. Luckily since the coils in our magnetic tweezers generate a relatively uniform B-field, they can act as a field extinguisher and no other specialised field modulator is necessary. The $x$ and $z$ components (see Figure 1 for a definition of axis) of the Earth's field can be cancelled by adding a constant current in the coils to generate a field equal but opposite to the Earth's field. The $y$ component adds to the rotation a constant angular offset from the $y = 0$ plane but it does not affect the rotation in the $y = 0$ plane, so the only effect it has on torque application is a constant reduction of the torque, which can be easily compensated by raising the coil current.

## 2.2 Bead rotation

We rotated the ferromagnetic bead over a range of frequencies both with and without the optical trap being turned on. When the optical trap is switched off, due to the remanence of the ferromagnetic beads, the beads tend to stick together in clumps of ~1-10 beads. We can take advantage of the asymmetry of the clumps to image the rotation (see Figure 4a). When the optical trapping field is present, only one bead is trapped but due to the imperfection of the bead surface, the rotation still creates an interference pattern in the back focal plane of the condenser, which is imaged onto a QPD to track the angular displacement of the trapped bead (Figure 4b). As the magnetic field is not separately measured, the degree of uniformity in the rotation of the bead is analysed and used to adjust the currents in each of the four coils to optimise the field rotation.

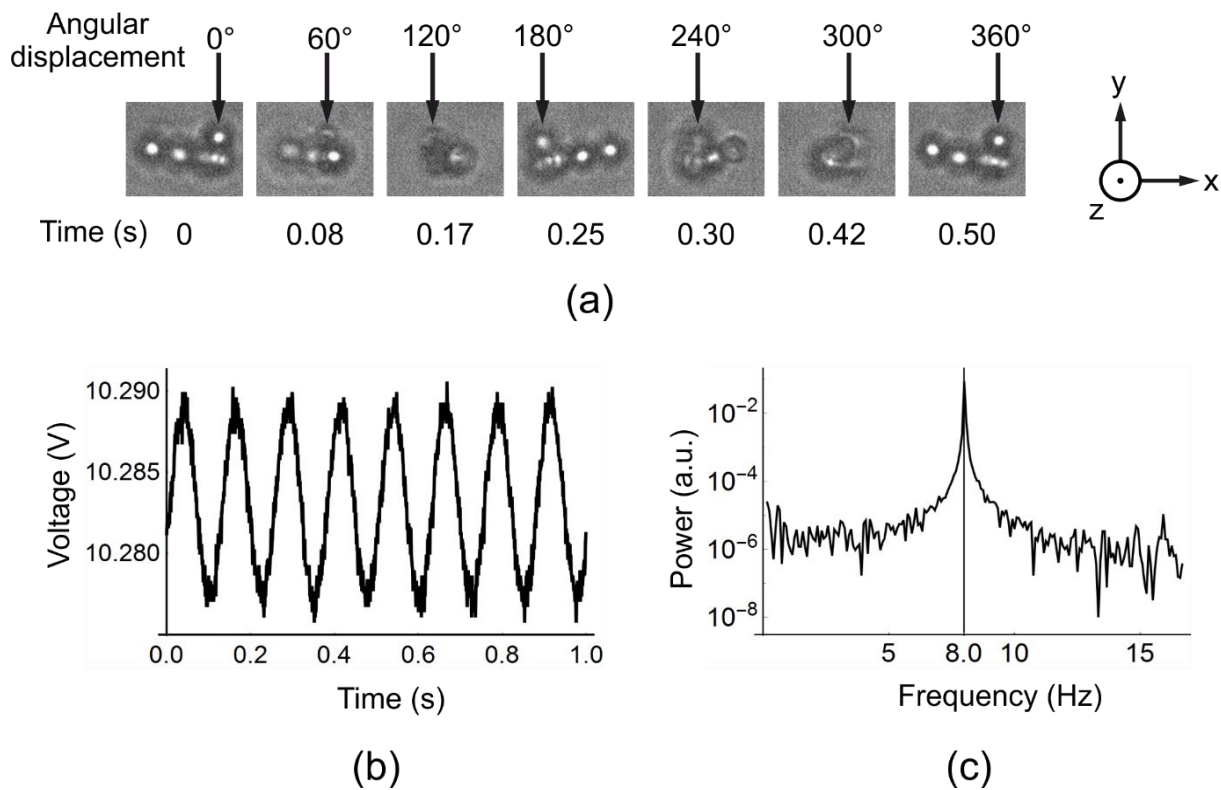

**Figure 4.** Rotation of ferromagnetic beads. In **(a)** the OT is not switched on so beads congregate and the inhomogeneity in the clump allows visual observation of rotation. The arrows at the top of each image point to the same bead throughout the rotation and the time at which each snapshot is taken is underneath each image. The rotation is at 2 Hz. The aggregation breaks up at higher frequencies due to viscous drag being larger on the peripheral beads. The fact that the structure is intact throughout rotations indicates that the field is uniform over that length scale. In **(b)** the OT is switched on and only one bead is trapped. The rotation frequency is set to be 8 Hz. The interference pattern is detected by a QPD. The periodic oscillation corresponding to the B-field rotational frequency is clearly seen. **(c)** plots the power spectrum of the oscillation, which shows a peak at 8.0 Hz.

## 3. Discussion

We present a novel combinatorial magnetic and optical tweezers in this paper. We also previously developed a super-resolution fluorescence imaging module, which is compatible with our tweezers module. The process of combining these technical capabilities into a single device is still on-going. When combined, it is our hope that the setup will have the ability to manipulate and image a single biological molecule simultaneously. This summation approach takes advantage of the tasks that each component is best at: magnetic tweezers are a robust and friendly technique to offer rotational control at defined angular velocity and at biologically relevant torque values; optical tweezers provide a versatile means to clamp and position the probe particles. The dynamics of DNA supercoiling of linear and circular double-stranded DNA have been explored at a range of ionic strengths, force and torque combinations and differing nucleotide base content *in vitro* and *in silico*. Direct fluorescence observation of the topological transformations will provide further insight into locational and dynamic information of such processes.

We have demonstrated the rotation of ferromagnetic beads at various frequencies and shown that due to the surface inhomogeneity of the beads we used, the rotation still can be detected as the interference patterns change in the condenser back focal plane even though they are macroscopically spherically symmetrical. The separation of linear and angular components maximises the versatility of both manipulation and measurement. Fluorescence imaging of a 48.5 kbp double stranded DNA molecule tethered to a magnetic bead on one end and coverslip surface on the other has been performed and described in a separate paper [34].

One limitation in optical trapping of magnetic beads, especially when an optically dense magnetic bead is trapped, is the absorption of laser energy by the bead and the surrounding medium, which leads to thermal damage in the sensitive biological sample, especially at the tethering point of attachment to the bead. Convection currents due to temperature gradients, changed local viscosity of the medium and increased Brownian movements are some other adverse effects of local heating. These heating effects have been estimated and measured [41]. Low force pulling experiments (for example, those in the physiologically relevant range of 0-10 pN) with simultaneous twisting with similar magneto-optical tweezers have shown minimal thermal effects on the system [26]. Also, efforts in reducing the local temperature rise, such as anti-reflection coating, have achieved significant success. Titania ($TiO_2$) coating reduces surface scattering and allows less than half of the laser power to achieve the same trapping stiffness [42]. A future phase of development in this project will be to investigate the use of surface modified beads to minimize heat absorption. But for now, even uncoated handles can be used in experiments where optical tweezers solely clamps the bead against diffusion, since minimum power is required for Langevin forces due to Brownian diffusion.

One great potential of our design is the improved temporal resolution in rotational degree of freedom afforded by the QPD back focal plane interferometry detection. But in the transverse MT configuration, the proximity of the bead to the flow-cell surface and the ~micron size of the bead both prevent the detection resolution from reaching its full potential as viscous drag limits the rotational bandwidth [26]. However, near-surface drag can be tackled with nanofabricated coverslip-surface patterns that bring the molecular tethering point far away from the coverslip surface, which we are currently developing. Also reducing the bead size to below 100 nm [43] or using 1 micron diameter microdiscs [44] are likely to significantly reduce drag, both aspects of which we will implement in future designs.

**Acknowledgements**

The authors thank John Dawson (University of York) and Ren Lim (University of Oxford) for discussions regarding current amplification for magnetic tweezers, Nynke Dekker (Delft University of Technology, The Netherlands) for discussions regarding magnetic bead choices and magnetic tweezers design, and Erik Hedlund (University of York) for technical support in software control of the lasers and the AOD. This work was funded by the Biological Physical Sciences Institute (BPSI) at the University of York.  MCL was supported by a Royal Society University Research Fellowship.

**Conflict of Interest**

The authors declare no conflict of interest.

# Magneto-optical tweezers with super-resolution fluorescence microscopy

**Zhaokun Zhou**[a, 1]**, Helen Miller**[a, 2]**, Adam J. M. Wollman**[a, 3] **and Mark C. Leake**[a, *]

[a] Biological Physical Sciences Institute (BPSI), Departments of Physics and Biology, University of York, York YO10 5DD, United Kingdom

[1] E-Mail: zz840@york.ac.uk;
[2] E-Mail: hlm531@york.ac.uk;
[3] E-Mail: adam.wollman@york.ac.uk;

[*] Author to whom correspondence should be addressed; E-Mail: mark.leake@york.ac.uk; Tel.: +44 (0)1904 322697 (Physics office); +44 (0)1904 328566 (Biology office).

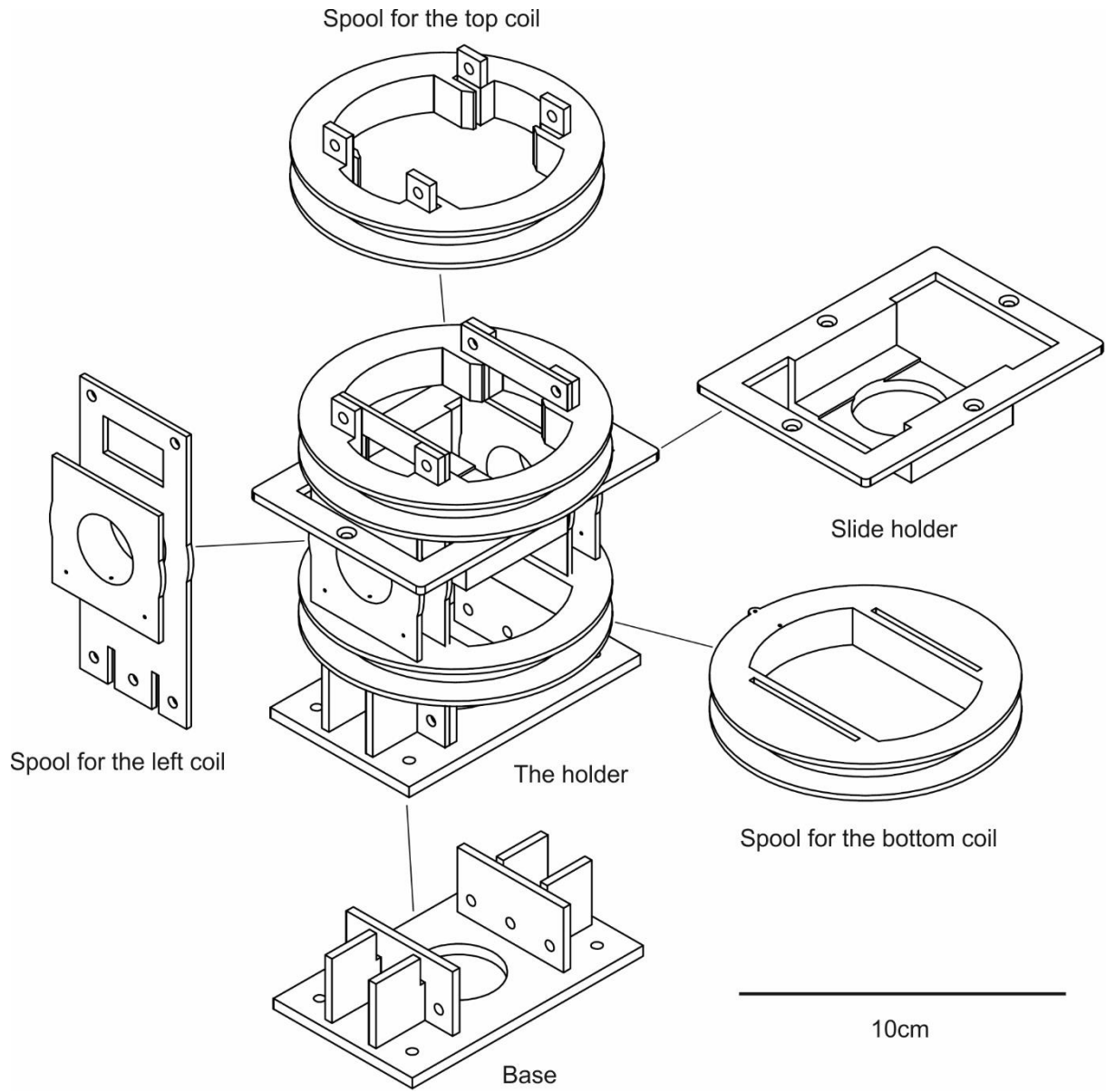

**Supplementary Figure 1.** Schematic of the bespoke holder of the electromagnetic coils and each individual component. The slide holder is mounted on the *xyz* nanostage and does not touch the other parts of the holder so it can move independently. The base provides the mechanism for the holder to be mounted on the microscope. The spool for the left coil slots into the top spool, the bottom spool and the base, holding the entire structure together. Note that the spool for the right coil is identical (but facing the opposite side) to this part so the right spool is not shown in the sketch. Readers can modify the design to suit their own experimental setups.

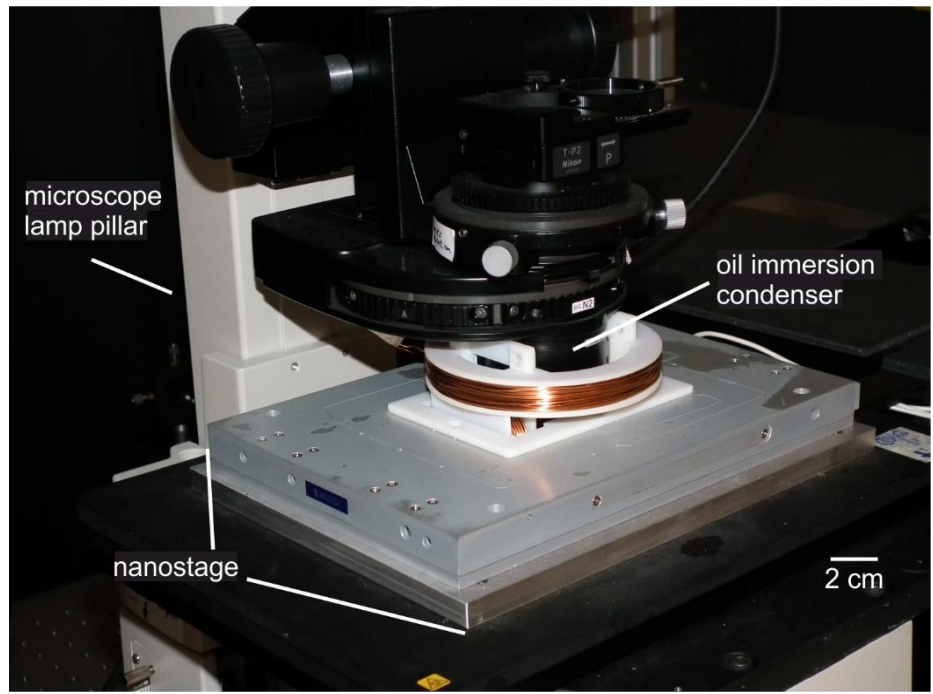

(a)

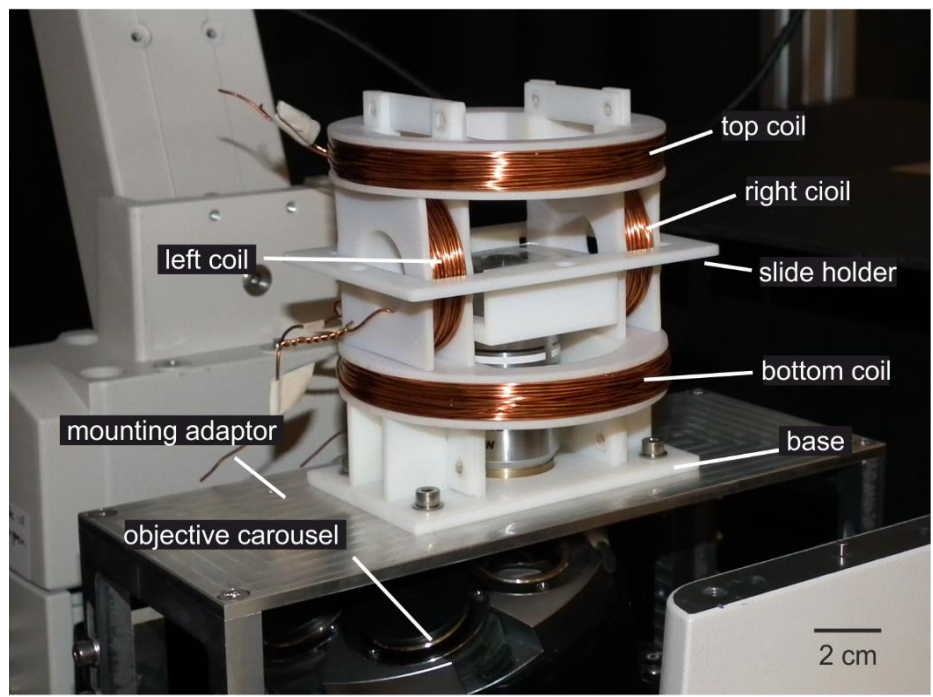

(b)

**Supplementary Figure 2.** Photos of the magnetic tweezers featuring the mounting mechanism to mount the MT onto the microscope so it is stationary relative to the objective lens, the 3D printed spools on which the enamel sheathed copper wires are wound and a bespoke sample holder that features a narrow tray to make room for the coils on the left and right. **(a)** Shows how the condenser, the magnetic tweezers and the *xyz* nanostage all fit together. **(b)** Exposes the magnetic tweezers structure for visualisation purposes here.

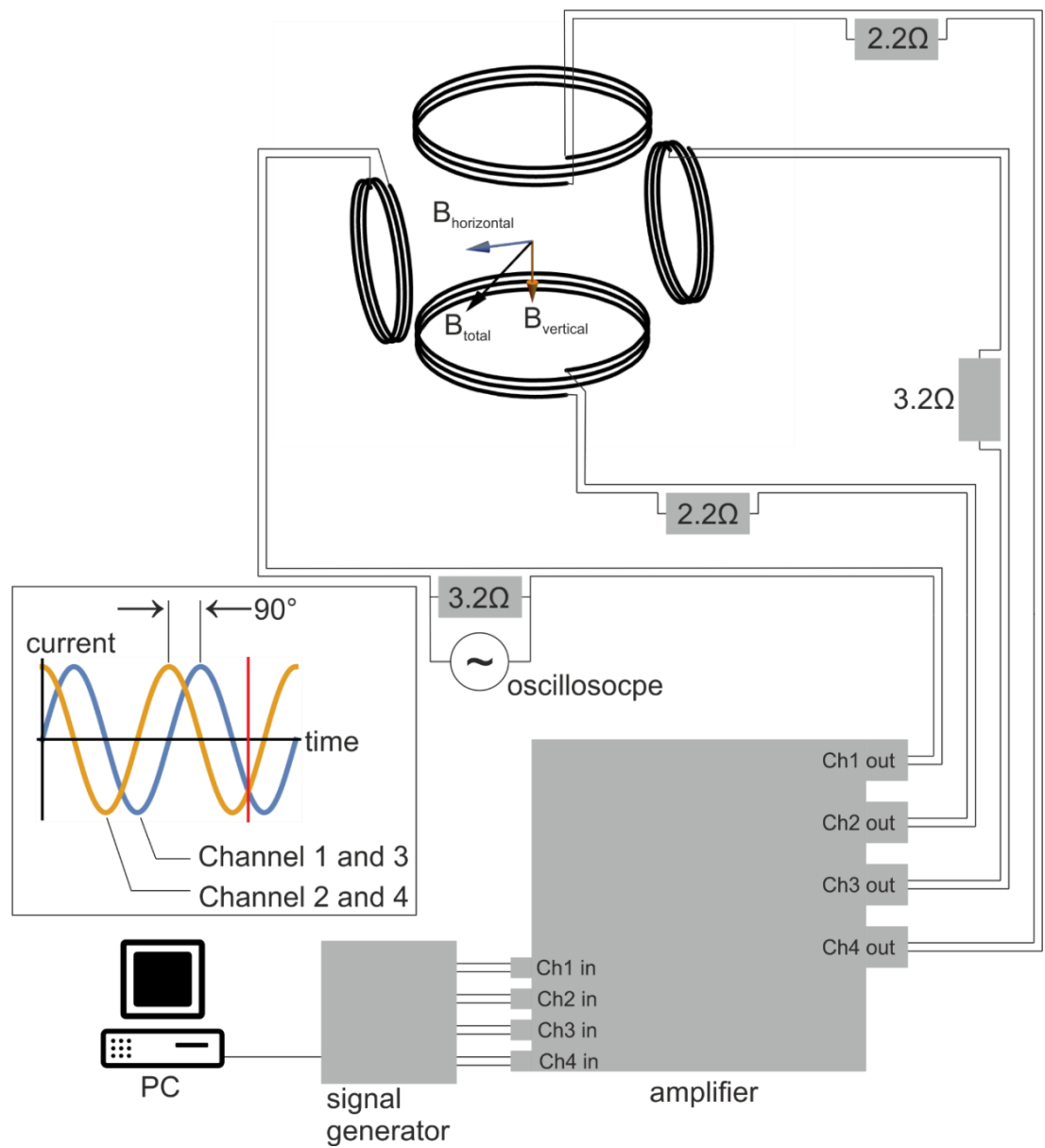

**Supplementary Figure 3.** Magnetic tweezers circuit diagram. Sinusoidal signal variations are synthesized with our home-made LabVIEW code. The PC sends the signal to a 4-channel signal generator (NI 9263, National Instruments), which is capable of creating AC voltages between ±10 V and at 100 kHz sample rate. This is then sent to a 4-channel car audio amplifier for high current output. Each channel has a resistor connected in series that adds to the impedance of the corresponding coil to bring the total load up to the rated output impedance of the amplifier. Also the voltage on each resistor is monitored with an oscilloscope (drawn only on Ch1 resistor) as a means to monitor the current in each coil. The inset on the left shows a typical current-time plot. Currents that are phased 90° apart are applied to each pair of coils. But the currents in both coils in either pair are the same. The red line in the inset corresponds to the B-field drawn in the centre of the coils. The orange arrow represents the combined field due to the vertical coil pair at the centre point, the blue arrow

represents that due to the horizontal pair and the black arrow represents the total resultant field due to all four coils.

**Heat dissipation calculation**

Each coil is modelled as a series of concentric rings for ease of calculation. The rings stack up in 10 layers. Each layer has a unique diameter and each layer has 10 rings (except the outermost layer of the small coil, which has 5, due to the fact that small coils only have room for 95 windings). The big coil is treated as 100 rings of diameters ranging from 38.5 mm to 47.5 mm; the small coil 95 rings from 14.5 mm to 23.5 mm.

The resistance of copper $\rho = 1.68 \times 10^{-8} \Omega$ m. The SWG 20 wire has diameter 0.914 mm, giving a cross-sectional area $A = 6.56 \times 10^{-7} m^2$. Thus the resistance of a ring is:

$$R = \rho \frac{l}{A} = 1.68 \times 10^{-8} \times \frac{2 \times 3.14 \times r}{6.56 \times 10^{-7}} = 0.161 \, r$$

where $r$ is the radius of the ring and SI units are used throughout.

The total resistance of the big coils is found by summing the resistance of each ring ($r$ takes values between 38.5 mm and 47.5 mm inclusive in steps of 1 mm):

$$10 \times \sum_r 0.161 \, r = 0.692 \, \Omega$$

And small ring ($r$ takes values between 14.5 mm and 22.5 mm inclusive in steps of 1 mm):

$$10 \times \sum_r 0.161 \, r + 5 \times 0.161 \times 23.5 = 0.287 \, \Omega$$

So the total resistance is:

$$R = 2 \times (0.692 + 0.287) = 1.96 \, \Omega$$

The root-mean-square current in the coils is

$$I = \sqrt{\frac{1}{T} \int_0^T [I_0 \text{Sin}(\omega t + \varphi)]^2 dt} = \frac{I_0}{\sqrt{2}}$$

where $I_0$ is the amplitude of the current, $\omega$ angular frequency, $\varphi$ angular offset, $t$ time and $T$ period. For $I_0 = 0.1$ A (a typical operating current), power

$$P = I^2 R = 20 \text{ mW}$$

Next we calculate heat capacity. The total length of all the coils is 76.4 m. Again the concentric ring model is assumed to obtain this value.

$$Capacity = V \times C = 76.4 \times 6.56 \times 10^{-7} \times 3.45 \times 10^6 = 172.9 \text{ J K}^{-1}$$

where $V$ is the volume of the coils and C is the isobaric volumetric heat capacity of copper. To raise the temperature by 0.1 °C, it takes a minimum of

$$t = \frac{E}{P} = \frac{172.9 \times 0.1}{20 \times 10^{-3}} = 8.6 \times 10^2 \text{s} = 15 \text{ min}$$

And this is in complete negligence of heat dissipation from the coils. In practice the temperature rise will be much slower.

**Dissipation *vs* wire thickness calculation**

Here we calculate the dependence of the rate of Joule heat generation on the thickness of copper wires that make up the magnetic tweezers. All other variables are held constant, such as the B-field generated and the space available in the spools for the wire winding. Also we neglect the thickness of the enamel wrapping of the wires and the skin effect.

The B-field is linearly proportional to the current, *I*, and the number of turns, *n*:

$$B \propto I \cdot n \tag{1}$$

which gives

$$I \propto \frac{B}{n} \tag{2}$$

The resistance of the wire, *R*, depends on the cross-sectional area, *A*, and the length, *l*, of the wire according to the following relationship:

$$R \propto \frac{l}{A} \tag{3}$$

Since $l \propto n$ and $A \propto \frac{1}{n}$ (spool space is fixed so the more turns there are, the thinner the wire needs to be), equation (3) can be written in terms of *n*:

$$R \propto \frac{n}{\frac{1}{n}} = n^2 \tag{4}$$

The equation for power dissipation is

$$P = I^2 R \tag{5}$$

Substituting (2) and (4) into (5);

$$P \propto \left(\frac{B}{n}\right)^2 n^2 = B^2 \tag{6}$$

The cross-section of the wire cancels out so Joule heating does not depend on the thickness of the wires.